\documentclass[referee]{aa} % for a referee version
\usepackage{graphicx}
\usepackage{bmpsize}
\usepackage{natbib}
\usepackage{amsmath}
\usepackage{wasysym}
\usepackage{txfonts}
\usepackage{soul}
\def\Msun{\hbox{$M_{\odot}$}}             %Msun
\def\Lsun{\hbox{$L_{\odot}$}}	%Lsun
\def\mic{\hbox{$\mu$m}}
\def\dgr{\hbox{$^{\circ}$ }}
\newcommand*\xoverline[2][0.75]{%
    \sbox{\myboxA}{$\m@th#2$}%
    \setbox\myboxB\null% Phantom box
    \ht\myboxB=\ht\myboxA%
    \dp\myboxB=\dp\myboxA%
    \wd\myboxB=#1\wd\myboxA% Scale phantom
    \sbox\myboxB{$\m@th\overline{\copy\myboxB}$}%  Overlined phantom
    \setlength\mylenA{\the\wd\myboxA}%   calc width diff
    \addtolength\mylenA{-\the\wd\myboxB}%
    \ifdim\wd\myboxB<\wd\myboxA%
       \rlap{\hskip 0.5\mylenA\usebox\myboxB}{\usebox\myboxA}%
    \else
        \hskip -0.5\mylenA\rlap{\usebox\myboxA}{\hskip 0.5\mylenA\usebox\myboxB}%
    \fi}

\begin{document} 
	\hyphenation{MontAGN}
	\title{Stellar cusp and warm dust   at the heart of NGC 1068 }
	\titlerunning{Stellar cusp and warm dust   at the heart of NGC 1068}
	
	\subtitle{}
	
	\author{ D. Rouan\inst{1}\thanks{\email{daniel.rouan@obspm.fr} }
		\and 
			L. Grosset\inst{1}
		\and 
			D. Gratadour\inst{1}\thanks{Based on observations collected at the European Southern Observatory, Paranal, Chile, through observing proposals 60.A-9361(A) and 097.B-0840(A).} 
			}
   \institute{
   			  LESIA, Observatoire de Paris, PSL Research University, CNRS, Sorbonne Universit\'es, UPMC Univ. Paris 06, Univ. Paris Diderot, Sorbonne Paris Cit\'e, 5 place Jules Janssen, 92190 Meudon, France
              }

   \date{Received , 24/10/2018; accepted , 11/01/2018}

% \abstract{}{}{}{}{} 
% 5 {} token are mandatory
 
  \abstract  % context heading (optional)
  {}
  % aims heading (mandatory)}
{
Establishing  precisely how  stars and interstellar medium  distribute within the central 100 pc area around an AGN, down to the pc scale, is key for understanding    the late stages of transfer of matter onto the accretion disc.
  }
   { Using AO-assisted (SPHERE-VLT) near-IR images in H and Ks and narrow-band of the Seyfert 2 galaxy NGC 1068 we  analyse the radial distribution of brightness in the central $r  < 100 $~pc  area down to the pc scale. The median-averaged radial profiles are fitted by a cusp (power-law) plus a central point-source. A simple radiative transfer model is also used to interpret  the data.}
   {We find that the fit  of radial brightness  profiles beyond 10pc  is done quite precisely  at Ks by a cusp of exponent -2.0  plus a central point-source and by a cusp of exponent -1.2 at  H. The difference of exponents between H and Ks can be  explained by differential extinction, provided that the distribution of dust  is itself  cuspy, with an exponent -1.0. However, the derived stellar density is   found to follow a $r^{-4}$ cusp, which is much steeper than any other cusp theoretically predicted around a massive black hole, or  observed,  in the center of early- or late-type galaxies or in mergers.  Introducing a segregation in the stellar population with a central excess of giant stars leads to a somewhat less steep exponent, however the de-redenned luminosity of the stellar cusp, as well as the mass of dust and gas  appear all much too high to be realistic.  
   An alternative scenario, where the Ks profile is well fitted by a combination of radiation from a stellar cusp identical to the H one and of thermal emission of warm/hot dust heated by the central engine, appears much more satisfactory. 
    We identify the  central point-like source with the very hot dust at the internal wall of the putative torus and derive an intrinsic luminosity that requires  a central  extinction $A_{K} \approx $ 8,  a value consistent with  predictions by several models of torus.  }
   {
  }


   \maketitle
%
%-------------------------------------------------------------------

%%-----------------------------------------------------------------

\section{Introduction}

In an Active Galactic Nucleus (AGN hereafter),  considered here as the very central region encompassing the stellar cluster and a  dense molecular core, the feeding of  the Central Engine  (CE in the following),  i.e. the accretion disk,  requires a matter accretion rate  from 0.3  in Seyfert \citep{Bian03} up to 10 M$_{\astrosun}~ yr^{-1}$ \citep{Hopkins12} for the most luminous QSO. How is achieved the very final transfer of matter from kpc scale to the sub-parsec size of the accretion disc is still a matter of debate since the mechanism of loss of angular momentum at   scale  below 100 pc is not well established (see \cite{Goodman03} for a review). Neither is clear any relationship between this feeding mechanism and the likely existence of an obscuring torus,  the paradigm proposed by  \cite{Antonucci1985} years ago to interpret the Seyfert 1/2 duality, and that still remains valid in the light of recent ALMA or SPHERE polaro-imaging observations (\citealt{Garcia-Burillo16,Gratadour2015}). One of the most difficult problems for hydrodynamical models is to take into account a huge range of scales, from the 10 kpc one of spiral arms down to the 0.1 pc of the  accretion disc, while integrating all possible processes such as cooling, star formation, self-gravity of the gas which are all  highly non-linear.  

Recently \cite{Hopkins12}, using results from their multi-scale hydrodynamical model (\citealt{Hopkins10a, Hopkins10b, Hopkins11}), proposed  that  the formation, through instabilities, of lopsided  eccentric gas+stellar discs of size of about 10 pc would be the dominant cause for angular momentum transport below this scale. The asymmetry of the stellar disc  would be the main cause of the strong torque exerted on the gas.   The model predicts also  that surface densities of gas and stars  both  follow  cuspy distributions.

 Performing observations in vis-IR that could test the validity of this proposal, especially evidencing the stellar disk, is not obvious because of several limits: {\it i})  we are dealing with scales which are largely subarcsec (1pc = 20 mas at 10 Mpc) ; {\it ii})  the central region is very obscured by the huge concentration of dust ; {\it iii}) the  bright dazzling AGN is totally dominating the flux, making difficult the detection, in its immediate neighbourhood,  of fainter structures  associated with  warm dust emission or stellar population. Of course millimetric interferometry, especially using ALMA, could solve the two first issues, but it is not  suited to probe stellar emission. On the other hand, the combination of adaptive optics on large  telescopes and imaging in the near-IR gives the opportunity to tackle the problem, at least on nearby AGNs. This is the  goal of this paper where we present an analysis of Adaptive Optics (AO) images of NGC 1068 in K$_{s}$-band and H, as well as in several narrow-band filters, obtained with SPHERE on the VLT.

NGC 1068 is a  Seyfert 2 galaxy which has  been the  target for many studies about AGN since it is one of the closest active nucleus (14.4 Mpc, following  \citealt{Bland-Hawthorn1997}) and therefore one of the brightest.   One consequence   is that the nucleus is bright enough to be used as the guide source for  the wavefront sensor of the AO system, allowing to obtain high angular  resolution images in the  near-Infrared (NIR), as e.g.   with PUEO-CFHT \citep{Gratadour2003}, NAOS-VLT (\citealt{Rouan2004,Gratadour2006,Prieto05}), SINFONI-VLT \citep{Davies07,Muller09},  or  more recently using SPHERE-VLT (\citealt{Gratadour2015}).

The later observations  were conducted during the SPHERE Science Verification (SV) program in December 2014. The H-band and K$_{s}$-band broad band images, showed a polarisation angle map with a clear centro-symmetric pattern, tracing both  parts of the ionisation bicone. Even more important, they featured a central non centro-symmetric pattern   approximately  60 pc $\times$ 20 pc wide, with aligned polarisation, that was interpreted as the trace of the outer envelope of the torus, revealed through a double scattering process (\citealt{Gratadour2015, Grosset18}). It is this set of images, complemented by images obtained in narrow-band filters in September 2016, still with SPHERE,  that we re-processed,   but without using the polarisation information. In this paper we concentrate on the  radial distribution of brightness.

The paper is organised as follows: the data processing and radial profile fitting is presented in section 2; the discussion on the power-law radial profiles  at H-band and K$_{s}$-band and their interpretation  in terms of a reddened cuspy stellar distributions through a simple transfer model is done in section 3, after some discussion about various classes of cusps observed in center  of galaxies.  In section 4, we discuss an alternative and more acceptable model where warm/hot dust heated by the central engine is mixed to stellar emission and finally,  we examine the  central point-like source in section 5 as well as  the constraints brought on the dust opacity by its luminosity, before concluding in section 6.

\section{Data set and data processing}
\label{sec-data}
\begin{figure}[ht!]
 \centering    
 \begin{tabular}{ll}
 \includegraphics[width=0.225\textwidth,clip]{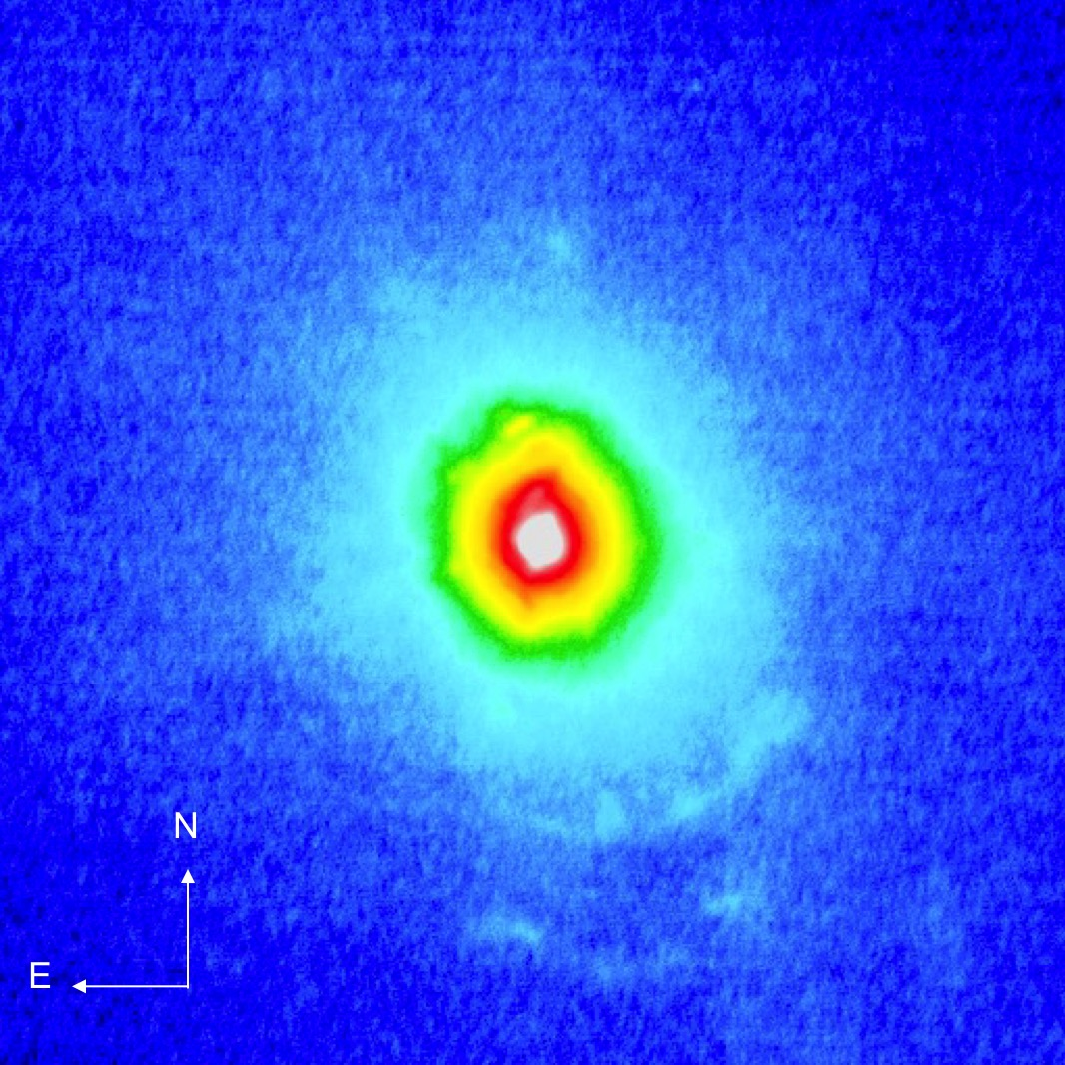} 
 &
 \includegraphics[width=0.225\textwidth,clip]{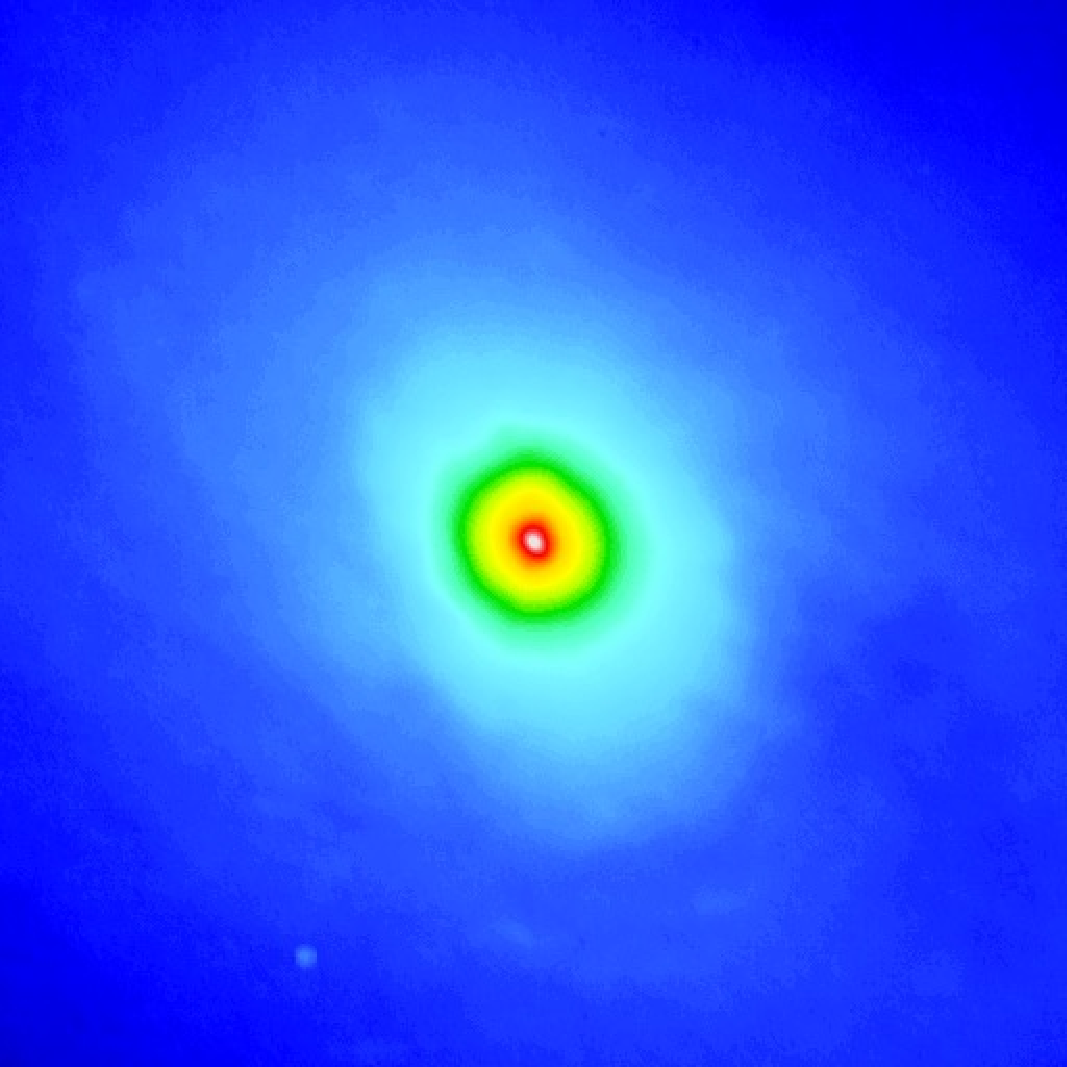}

\end{tabular}
  \caption{K$_{s}$-band (left) and H-band (right) images of the central 438 $\times$ 438 pc of NGC 1068, obtained with SPHERE. 1 arcsec is 70 pc.  The intensity scaling corresponds to a Asinh law. For the K$_{s}$-band  image, the flux at peak (white area)  is 2.3 $10^{-3}$ Jy/pixel and in the light blue area it is $\approx 6.0 10^{-6}$ Jy/pixel. For the H-band  image, the flux at peak (white)  is  $2.0 10^{-4}$ Jy/pixel and in the light blue area it is $\approx 7.5 10^{-6}$ Jy/pixel. }
  \label{F2}
\end{figure}

 The  K$_{s}$-band and H-band images obtained with SPHERE in December 2014 are shown in Fig. \ref{F2}, using a Asinh scaling law for the intensity, that enhances the low flux levels.  Details about observations and data reduction have been published in  \cite{Gratadour2015}. Pixel size is 12.25 mas, corresponding to 0.855 pc assuming a distance of 14.4 Mpc. The field of view is 6.27 $\times$ 6.27 $arcsec^{2}$, or 438  $\times$ 438 pc$^{2}$, and the angular resolution, as measured on the PSF,  is  58 mas and 53 mas, respectively in K$_{s}$ and in H.

On the H-band image one can note  the elongated disc, slightly asymmetric  with respect to the CE,  well defined by a strong peak. The disk size is  about 135 pc $\times$ 180 pc, choosing as the limit, the distance where the slope of the radial distribution flattens.  Assuming that the corresponding iso-contour   
delineates  the disk, the barycenter of this contour is offset by 6 pixels ( 5.1 pc) to the West and 1 pixel  ( .9 pc) to the North with respect to the central peak. 

As regards the K$_{s}$-band image, the asymmetry is less pronounced and, for instance, the center of gravity of
the contour delineating about the same area is only shifted of 2.5 pixels (2.1 pc) and 1.5 pixels (1.3 pc) respectively to West and North. 

Now, an important question is: what is the source of emission at H-band and K$_{s}$-band? The answer is not trivial,
especially concerning the K$_{s}$-band. Indeed we expect an important  contribution from stellar emission, but we also expect a significant contribution from very hot dust, close to sublimation temperature, arising from the internal radius of the likely torus. For instance, \cite{Gratadour2003} showed clearly that the steeper slope of the near-IR spectrum when approaching the central peak is well explained by such very hot dust. In the following, our main concern will be to discriminate between the respective contribution of the two sources of radiation.

Let's first examine  the medium scale. We present  in Fig. \ref{F3a}  the median brightness profile in H-band  extending up to 430 pc, i.e. well beyond the molecular ring of r $\approx$ 200 pc {detected by \cite{Muller09} and confirmed by ALMA \citep{Garcia-Burillo2014}. On the same figure, two fits are represented: one as a power-law for the most internal part (r $\leq 110 pc$) and an exponential one for the external part, up to r = 430 pc. Both fits are fairly good and, assuming  that's essentially the stellar component which is seen at H, one  concludes that when going inward, a  disk distribution (here of scale length  $R_{c} = 410 $pc)  is superseded by a cuspy (i.e. power-law) distribution of exponent 1.2 for $r \leq 110 pc$. 

 \begin{figure}[ht!]
 \centering    
 \includegraphics[width=0.45\textwidth,clip]{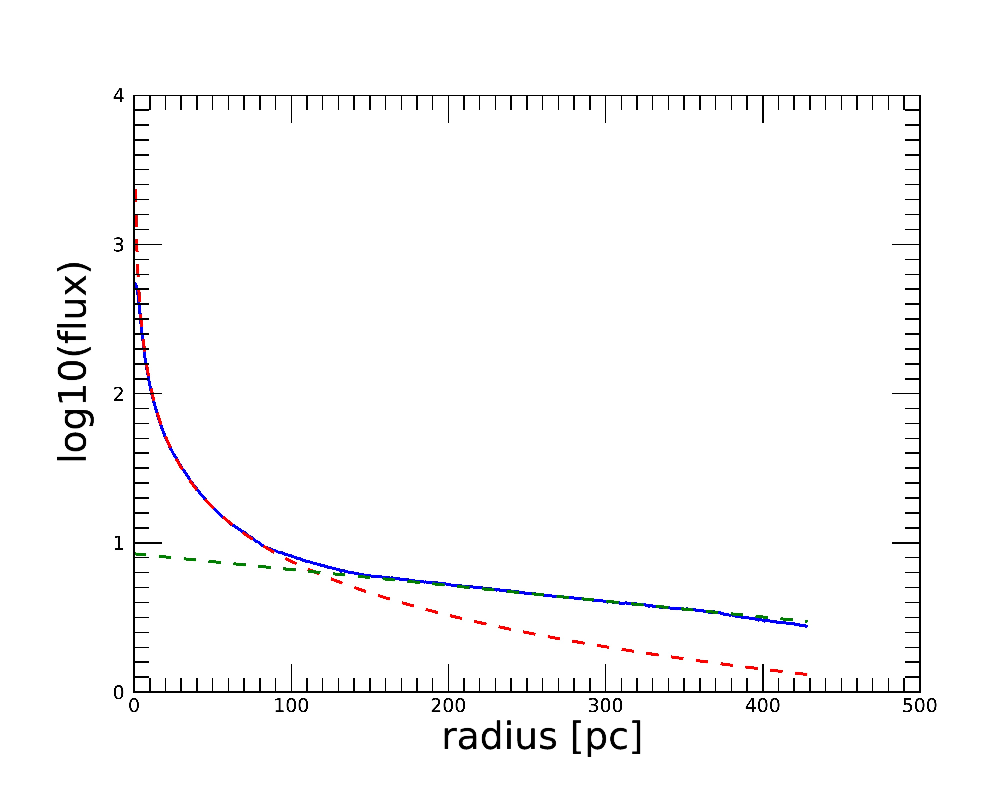} 

  \caption{Blue solid line: brightness profile in H-band up to 430 pc, estimated by performing a median of the pixels on successive circles centred on the central peak. Red dash line: power-law profile $r^{-1.2}$ fitted on the central distribution. Green dash line: exponential profile fitted on the external observed distribution. }
  \label{F3a}
\end{figure}

 In Fig. \ref{F3b} we plot the median radial  brightness profile around the central peak, measured on the K$_{s}$-band and  H-band images, together with the PSF profile at K$_{s}$-band measured on the nearby reference star BD-00413 whose visible magnitude is close from the one of the nucleus of NGC 1068, so that the AO performances can be considered as very similar.  One can note several features. First the H-band profile is almost perfectly fitted by a power-law of exponent -1.2; we will come back in the next section on the meaning of this behaviour. Second, we note that the K$_{s}$-band profile aisles, beyond  a radius of about 17 pc, is also well fitted by a power-law, but of exponent  -2.0. Interestingly, a fairly good fit to the whole data set is provided by a combination of a this power-law profile  truncated at r $< 4$pc  plus the PSF profile properly weighted. This appears as the blue curve  in Fig. \ref{F3b} superimposed on the K$_{s}$-band actual profile. In particular, we note that the small bumps on the K$_{s}$-band profile, close to the peak, are well fitted by the first and even the second Airy rings well appearing on the PSF  profile. The straightforward interpretation of this fit is that the K$_{s}$-band profile results from the contribution of a central point source and of a smoother distribution that is directly related to the H-band emission. We have tested other combinations by convolving the PSF with  a Gaussian of various widths, but beyond a FWHM of 2 pixels (1.7 pc), the fit degrades. The central  source is thus practically point like and in any case its radius is not significantly larger than 0.8 -- 1pc.

 \begin{figure}[ht!]
 \centering    
 \includegraphics[width=0.45\textwidth,clip]{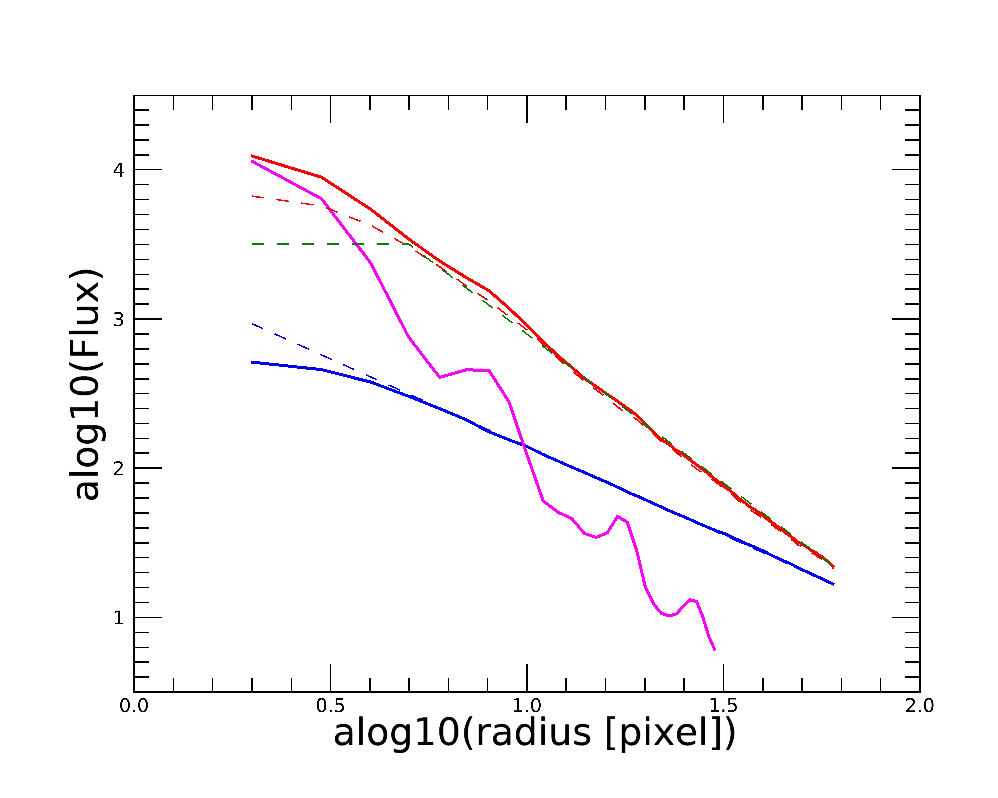} 
% \includegraphics[natwidth=7cm]{Hopkins11f2cusp.pdf} 

   %\caption{Simplified sketch of unified model of AGN structures.}
  \caption{Median radial  brightness profile around the central peak. Solid red :  measured at K$_{s}$; purple : PSF  at K$_{s}$; red dash:  difference between K$_{s}$-band and weighted PSF; green : truncated  power-law profile in $r^{-2}$ ; solid blue : H-band profile;  blue dash: power-law profile in $r^{-1.2}$.}
  %%%%%\`u A REMPLACER PAR UNE FIGURE PLUS EXPLICITE EN LOI DE PUISSANCE%
  \label{F3b}
\end{figure}

We will come back in section \ref{Sec_PointSource} on the interpretation of this important result that there is  a quasi-unresolved central point source in K$_{s}$-band with no obvious counterpart in H. We concentrate in the next two sections on the power-law, or cuspy, component. 

In Sept 2016 we also obtained images with SPHERE in three narrow-band filters noted cont--H (1.558~\mic, $\Delta \lambda$ = 0.0245~\mic  ), cont--K1 (2.103 ~\mic, $\Delta \lambda$ = 0.033~\mic ) and cont--K2 (2.267 \mic, $\Delta \lambda$ = 0.034~\mic ). The detail of the data reduction will be described in \cite{Grosset19}, a paper dedicated to the analysis of polarisation vs wavelength, but can also be found in  \cite{Grosset17}.  The data processing is essentially the same as the one used for the H-band and K$_{s}$ observations of Dec. 2014.  

We plot on Fig \ref{HK1K2K} the median averaged profiles at the four wavelengths on a log-log scale. One notices {\it i}) the fairly linear behaviour of all profiles from 4 to 100 pixels (3.5 to 85 pc), as illustrated by the superimposed dash lines   and {\it ii})  the absolute value of the slope   increasing  with wavelength, from H-band to K$_{s}$-band. The best fitted power-laws for the range 5 to 60 pc have exponents     -1.20, -1.36,  -1.95, -2.11 and  -2.03, for respectively H-band, cont--H,  cont--K1, cont--K2 and K$_{s}$-band. 

 \begin{figure}[ht!]
 \centering    
 \includegraphics[width=0.45\textwidth,clip]{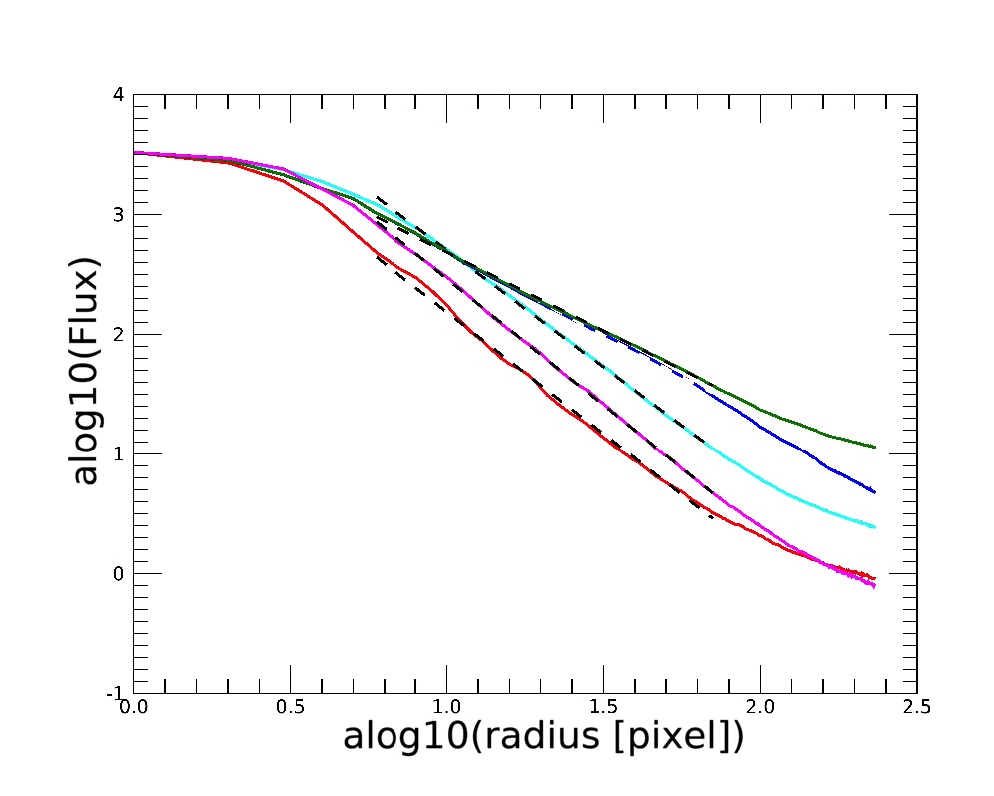} 
  \caption{Median radial  brightness profile around the central peak, for the  cont--H (blue), H-band  (green), cont--K1 (cyan),  cont--K2 (magenta) and K$_{s}$-band (red) images. The log-log scale allows to identify a power-law behaviour for all four bands. cont--K1, cont--K2, and K$_{s}$-band fluxes have been scaled on the H-band flux at the center. The dashed line on each curve is the best linear fit for the range 5 to 60 pc.  }
  \label{HK1K2K}
\end{figure}

There are several possible origins for the power-law emission, and we'll examine  different cases  in the next sections,  however, the monotonic variation of the slope with respect to  wavelength on a sample of 5 different  filters  is a strong indication that the cause of this variation is rather a continuously variable phenomenon  than some spectral features such as emission lines. With this hypothesis in mind  and taking into account the fact that the SNR is significantly lower in narrow-band images, we have chosen 
in the following to consider only the H-band and the K$_{s}$-band bands in our analysis, so that  thanks to the 
broad range in wavelength coverage, any effect  depending on wavelength will  be clearly enhanced.

The first hypothesis, which could be well in line with the quite similar behaviour of  the H-band and K$_{s}$-band   radial brightness profiles and  corresponds to a situation often observed   in central region of galaxies \citep{Boker08},  is that in both bands  we sense the emission from a stellar cusp, also called nuclear stellar cluster. 
The second case, which is generally invoked in the case of NGC 1068 \citep{Thatte97,Gratadour2003} and other AGNs \citep{Leja18}, is that there  is warm to  hot dust, heated by the CE, and radiating in the infrared, especially at Ks.
Of course, some mixing  of this dust thermal radiation with the photospheric emission from a stellar cluster is likely, especially at H.    

\section{A reddened stellar cusp ?}

In first instance, we can  assume that the H-band power-law profile is largely due to the stellar emission. First, because of this quasi-perfect power-law behaviour which is very similar to what is observed in many nuclei of galaxies (see next section) while it could hardly be explained by  dust emission extending so far from the CE and, second, because previous spectroscopic observations  by \cite{Thatte97} have clearly identified  the 1.6 \mic ~3$\rightarrow$6  and the 2.29 \mic  ~0$\rightarrow$2  band-head CO feature in a
central core of FWHM  of size around 45 pc (0''.66). However,   \cite{Thatte97} estimate the contribution from stellar emission to be only 1/3 of the total within an aperture of 1'' radius. Note that the angular resolution they reached is only 0''.9, thus far below the SPHERE performance of 0''.06 in K$_{s}$-band (see  sect. \label{sec-data}).

Concerning the K$_{s}$-band profile, the contribution of hot dust from a compact  central source was clearly identified as early as 1987 \citep{Chelli87}.   \cite{Thatte97}, based on their 0''.4 resolution speckle observations estimated that essentially all the flux at K$_{s}$-band could be attributed to the point-like source. This is however not what we observe on the K$_{s}$-band profile of Fig. \ref{F2}, where the aisles extend far beyond those of the PSF.  On the other hand, the fact that the aisle beyond 17 pc (0''.24) are very well fitted  too by a power-law leads to the same reasoning as for the H-band, i.e. that it could correspond  to stellar emission.  If this hypothesis is correct, then it raises  the question: why are the exponents so different between K$_{s}$-band and H-band ? 

Another question  raised by this profile behaviour is: how the cusp characteristics compare with those of cusps actually found in early-type galaxies or in LIRGs / ULIRGs ?

As regards the size of the cusp, by integrating the flux in disks of increasing  diameter,  we have  
determined a characteristic  half-light radius of 99 pc (1.420'') and 17 pc (0.245'')   for  
the H-band and the K$_{s}$-band cusps respectively,  thus two very different scales, a fact that we should try to explain. 

Compared to the structures observed at millimetric wavelengths, either in molecular lines (CO, HCN) or in the continuum \citep{Garcia-Burillo2014,Garcia-Burillo2016,Imanishi2018}, we note that the extent of the cusp at H-band reaches the inner radius of the off-centered circumnuclear molecular disk. This means that we do not expect significant extra extinction from the disk itself, since the stellar cusp is almost entirely within its inner contour. 

\subsection{Emission lines from the ISM}

We'll consider in section \ref{Sec_extinction} the possible role of differential dust extinction to explain the log-log H-band vs K$_{s}$-band slope difference, but before, let's first examine  another potential explanation, if there is an extra contribution  to the K$_{s}$-band flux of lines emitted by the ISM,  that would be both significant and featuring some gradient.  One can think of two kinds of line emission:  either the emission of the (1-0) S(1) ro-vib line (2.1218 \mic) of some excited H2 , or the Br$_{\gamma}$ line  (2.1655 \mic) associated to the narrow- or even the broad-line region. The first case must be excluded since the  high angular resolution  map of the line, obtained by \cite{Muller09} with the spectro-imager SINFONI-VLT 
clearly shows that the line emission is confined in a compact region ( about 0.15'' diameter) at 0.15 '' to 
the South-East of the central engine.  Moreover, the same authors provided a spectrum of the region
0.3'' $\times$ 0.1'' centered on the nucleus that clearly shows that the contribution of the  (1-0) S(1) line
to the flux in the whole K$_{s}$-band band is only 5. $10^{-4}$. The same spectrum indicates that the Br$_{\gamma}$ line contribution to the same band is approximately 3. $10^{-3}$ and thus should also be excluded as an explanation of the  differences in profile slopes between H-band and Ks$_{s}$-band. A same conclusion can be reached considering the long slit spectrum (0.3 '' slit width)  in band K$_{s}$-band obtained by \cite{Gratadour2003}. 

\subsection{The variety of stellar cusps}
In the literature, the notion of central cusp seems to refer  to a variety of situations or contexts and we 
first try to make more explicit this variety. 

We can identify a first category, where  the  cusp is associated to some compact central stellar cluster. 
For instance, it corresponds to some {\it extra-light} compared to a S\'ersic  distribution  
extrapolated to the very center  in early-type galaxies \citep{Kormendy09,Lauer07}. The expression {\it 
nuclear  cluster} (NC) may also be  used, either for early or  late-type galaxies \citep{Boker04}. 
\cite{Lauer07} use the notion of {\it power-law galaxies} (with steep brightness increase) as opposed to 
core galaxies where the central brightness is nearly constant. \cite{Cote06} use the denomination {\it compact central nucleus}.  We consider in the following that all these denominations are representing the same kind 
of object, namely a central compact stellar cluster analog to a globular cluster, even if it differs from this 
class by a higher luminosity (see e.g.  \cite{Boker04}), typically 3.5 mag brighter. 
As stated by \cite{Boker08} NCs are common: the fraction of galaxies with an unambiguous NC detection is 75 \% in late-type (Scd-Sm) spirals \citep{Boker02}, 58\% in earlier type (S0-Sbc) spirals \citep{Balcells07}, and 70\% in spheroidal (E and S0) galaxies \citep{Graham03,Cote06}.The median 
characteristics that can be for instance derived from table 2 of \cite{Lauer07} is a half-light radius of 7 pc, 
within a range of 2 to 44 pc, and an exponent 
of 0.62, within a range between 0 and 1.1. \cite{Cote06} find for Virgo early-type galaxies a similar and 
consistent radius of 4.2 pc for a range extending to 62 pc. Those typical characteristics do not depend 
strongly on the type of  galaxy, considering for instance the study by \cite{Boker04} who find, for a 
sample of late-type galaxies, an histogram of radii    with a first peak  at 3 pc and an another one at 8 pc, 
the range extending to 30 pc. However,  it is well established  that faint ellipticals ($M_{V} \leq   20$)  
exhibit a more cuspy profile than massive ones   \citep{Kormendy09, Cote06}, a feature that likely 
translates into some different history of merging.

The second and distinct category of stellar cusps we have identified in the literature is more 
clearly related to recent or on-going starburst. We refer in particular to the study by \cite{Haan2013} of  a large 
sample of LIRGs and ULIRGs where HST images in near-IR were used to derive the characteristics of 
the nuclear stellar cusps they found in nearly all objects. 
We have analysed the data compiled by those authors concerning $\approx$ 80 
LIRGs and ULIRGs objects, a large fraction of which being mergers at various phases of the 
process. In most objects the authors found a cuspy profile in the nuclear region.  They  provide a table 
in which are compiled various properties for the selected objects, including radius 
and near-IR luminosity of the cusp, as well as the exponent of the cuspy brightness distribution.
Clearly the average characteristics are  distinct from those of the  class we discussed above:
we find a median half-light radius of 475 pc within  a range between 46 pc and 2.4 kpc, while the
exponent of the cusp has a median value of 0.80 within a range between 0.1 and 1.7. The cusp radius 
is remarkably  much larger than for the nuclear cluster of the  class identified above and the exponent is significantly larger and can reach high values. 

Finally the third category of cusps that is described in literature corresponds to the compact stellar 
population which is  distributed around a massive black hole in the center of a galaxy. Some theoretical works 
 have explored the dynamics of this peculiar system
(see e.g. a  review by \cite{Amaro-Seoane10}) but there are no well established observational facts,
with the notable exception of  the center of the Milky Way  \citep{Genzel03,Schodel17}.
Those NSC (Nuclear Stellar Clusters) cusps  are 
characterised by half-mass or half-light radii of 1-5 pc and  are in principle the result of dynamical effect, 
namely two-body relaxation through frequent encounters of stars or stellar BH in the strong gravitational 
field of a massive BH. Especially, \cite{Bahcall76}  established the first solid analytical expression of the 
stellar distribution around a massive BH and showed that it must be  cuspy with an exponent  1.75 (7/4). 
Refinements to the theory were brought later by numerous teams (for a review, see e.g. \cite{Preto10}), 
especially the fact that there must be a mass-segregation with different exponents according to the mass 
of the stellar objects, the most massive (stellar BH) having the largest exponent (1.8) and main sequence stars a lower value of typically 1.3 \citep{Freitag06}. Such exponents are thus clearly larger than the one found in the two previous classes we examined, and closer from the ones observed in NGC1068, but if we consider both the predicted small half-mass radius and the too large  relaxation time $T_{r}$ required to  reach to a cuspy distribution -- due to the large mass of the BH --, we'll no longer consider this   case for  NGC 1068.

To make the distinction between the  two classes more readable in the following, we will call Nuclear Cluster  Cusp (NCC)  the first one, Central Starburst Cusp  (CSC) the second one. Our goal is to try to find if there are clues allowing to identify the cusp we observe with one of those classes, or a combination.

The rather large exponent we found for the cusp in NGC 1068, as well as the rather wide half-light radius
are clearly ruling out the case where we would observe a mere NCC: such large exponents are never observed for this class. 

As regards the CSC class, NGC 1068 shares, at least for the H-band cusp, some characteristics, with an exponent and a radius that are indeed significantly different from the median values, but  within the observed ranges for this class. Nevertheless, the K$_{s}$-band
profile with its small half-light radius and large exponent could  weaken this statement, but we'll see in section \ref{hot_dust} that the dust emission cannot be neglected. However, there is  such a large dispersion of characteristics in the sample of \cite{Haan2013} that  the question deserves some closer examination.    

To make a step further,  starting from the data set compiled by \cite{Haan2013}, we have searched if some general law could relate the parameters of the cusp in CSC. The purpose is to see if NGC  1068 could obey such a general law,  even if it is not considered as a LIRGs. In  case where the answer is yes, this would be  a strong indication that the  cusp is rather  the result of a starburst evolution than of a dynamical effect around a massive BH. We first selected the 62 objects where the brightness distribution is indeed cuspy ($\gamma \ne 0$) and where the information on the three parameters: radius, near-IR luminosity  and exponent  of the cusp, was complete (e.g. not given as an upper or lower limit).  We looked for a fit of $\gamma$  as a linear combination of the log of the  near-IR cusp luminosity $L_{cusp}$ and the log of the cusp radius $ R_{cusp}$. We  found that the law 

\begin{equation}
 \gamma = 1.133\times [log10(L_{cusp} / R_{cusp}) - 6.63  ]
\label{Eq1} 
\end{equation}

relating the three parameters was rather robust, with a covariance between the actual value and the predicted one of .85. The robustness of the fit can be appreciated on Fig. \ref{Fig_correl}, where the  luminosity of each selected object is plotted against the one predicted by the law using the two other parameters.  
\begin{figure}[ht!]
 \centering    
 \begin{tabular}{ll}
 \includegraphics[width=0.50\textwidth,clip]{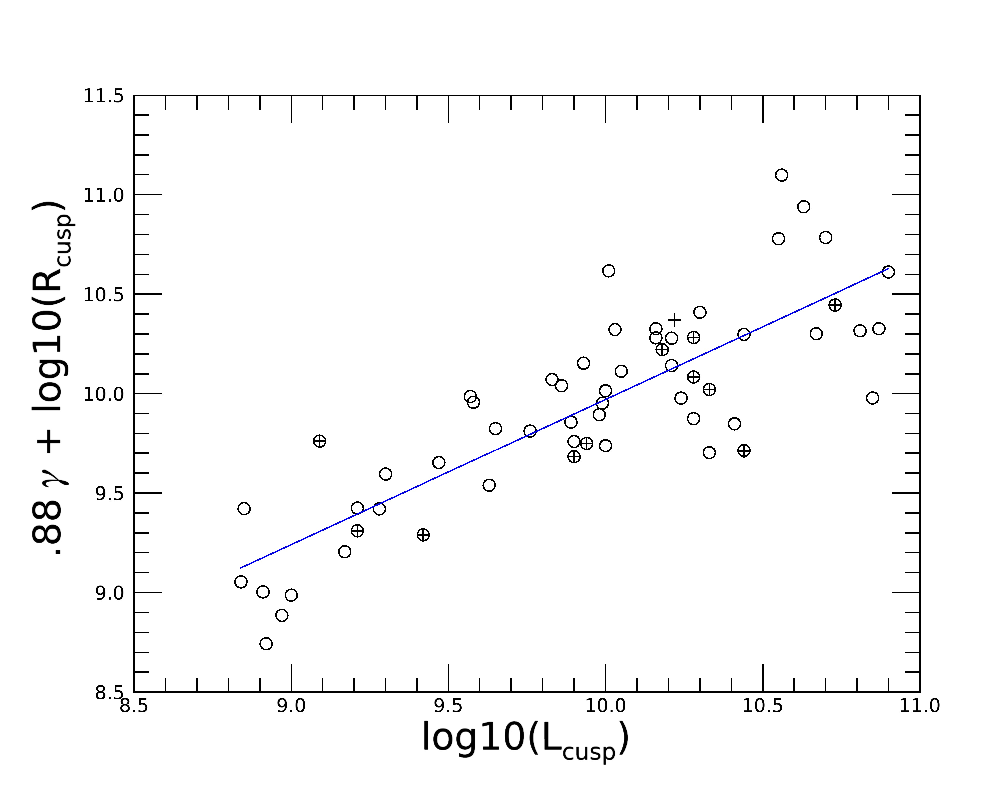} 

\end{tabular}
  \caption{Plot of the estimated cusp  luminosity  of a selection of the objects studied by \cite{Haan2013}, as predicted by the phenomenological law we found (see text), vs the actual cusp luminosity. The blue line is the linear regression. The cross indicates the objects which are not mergers or are in a wide system.}
  \label{Fig_correl}
\end{figure}

The straightforward interpretation of the law is that at a given luminosity, the smaller the radius of the cusp, the steeper the distribution and conversely, to a lower luminosity corresponds  a smoother cusp.  
Note that on Fig. \ref{Fig_correl}, we have  marked with an additional cross the objects which are not belonging to a merger  or are members of a system with widely separated galaxies (class 0 and 1 in \cite{Haan2013}): we cannot see any trend indicating that   those objects do not follow the general law we found. This is important since, as noted by \cite{Haan2013},  the fact that there is no obvious merging situation does not prevent a cusp to be present. 

In order to compare the characteristics of the NGC 1068 cusp with the derived law, we have to estimate the near-IR luminosity of the cusp. Because the obscuration is extremely high in the central region, we cannot  consider visible data as it was done in the previously mentioned studies of cusps in galaxies. 

To be consistent, we consider only the H-band data which are the less affected by possible dust emission, assuming that it 
represents  well the stellar population distribution. With a calibration  done with the star BD-00413,
observed the same night,  we evaluate a magnitude m$_{H}$ = 9.5 within an aperture of radius 100 pc. Assuming a 
stellar population with H-K$_{s}$-band $\approx$ 0  and an  age   between 4.0 $10^{8}$ and 6.0  $10^{9}$  yrs, the bolometric 
luminosity would be between 3.7 $10^{9}$ and 1.8 $10^{10}$ \Lsun, according to Fig. 5 of \cite{Thatte97}.   This is 
consistent with \cite{Davies07} who derives a bolometric luminosity of  stars  in the center of  7. $10^{9} \Lsun$. 
Using those estimates and the half-light radius 100 pc in Equ. \ref{Eq1}, we find a predicted exponent between 1.06 and 
1.84, a range that brackets the exponent  1.3 observed on the H-band brightness profile.   

This being said, we must emphasise the somewhat large uncertainty linked to the evaluation of the bolometric stellar 
luminosity. This rather large exponent of 1.3 is also partly consistent with \cite{Muller09} who  estimated  an enclosed 
stellar mass  varying linearly with the radius ( $M_{\star}(r) \propto r $), thus requiring a stellar density varying as $  r 
^{-2}$. The agreement with the empirical law we established,   suggests   that the cusp in NGC 1068 could be well 
related to some starburst activity. Indeed for \cite{Davies07} there are strong indications, based on Br$\gamma$ 
equivalent width, supernova rate and mass-to-light ratio,   that the stellar nucleus of NGC 1068 is the remnant of recent, 
but no longer active,  starburst  episodes. The authors estimate the age of the last episode to be 200-300 Myr.

 The only viable class to which the NGC 1068 cusp appears  to be related to is the CSC. We note that  \cite{Mushotzky14}, using {\it Herschel} data on a sample of X-ray selected AGNs,  reach the conclusion that most active galactic nuclei live in high star formation nuclear cusps. It remains however a question: if we do see mainly stellar emission, why are the exponents and half-light radii so different at H-band and K$_{s}$-band ? A first  tentative explanation  is that this difference is due to the effect of extinction by dust which may play an important role in this region. Another possibility is that there is a segregation in stellar 
 population with redder stars (giant) dominating towards the most central part.  A combination
 of both effects is, as well, possible. We examine those two cases in the next subsections. 
 
 \subsection{Pure differential dust extinction}
 \label{Sec_extinction}
 The larger dust absorption at H-band compared to K$_{s}$-band must smooth  more strongly the  stellar brightness as we approach the central peak, provided that the dust quantity increases inward.  To test such an effect,  we developed a rather simple  radiative transfer model, assuming that the  ISM  in which is embedded the cuspy distribution of stars follows as well a cuspy, i.e. power-law, distribution, but with possibly a different exponent.   Fig. \ref{Fsketch} depicts the geometry used.  
 \begin{figure}[ht!]
 \centering    

 \includegraphics[width=0.25\textwidth,clip]{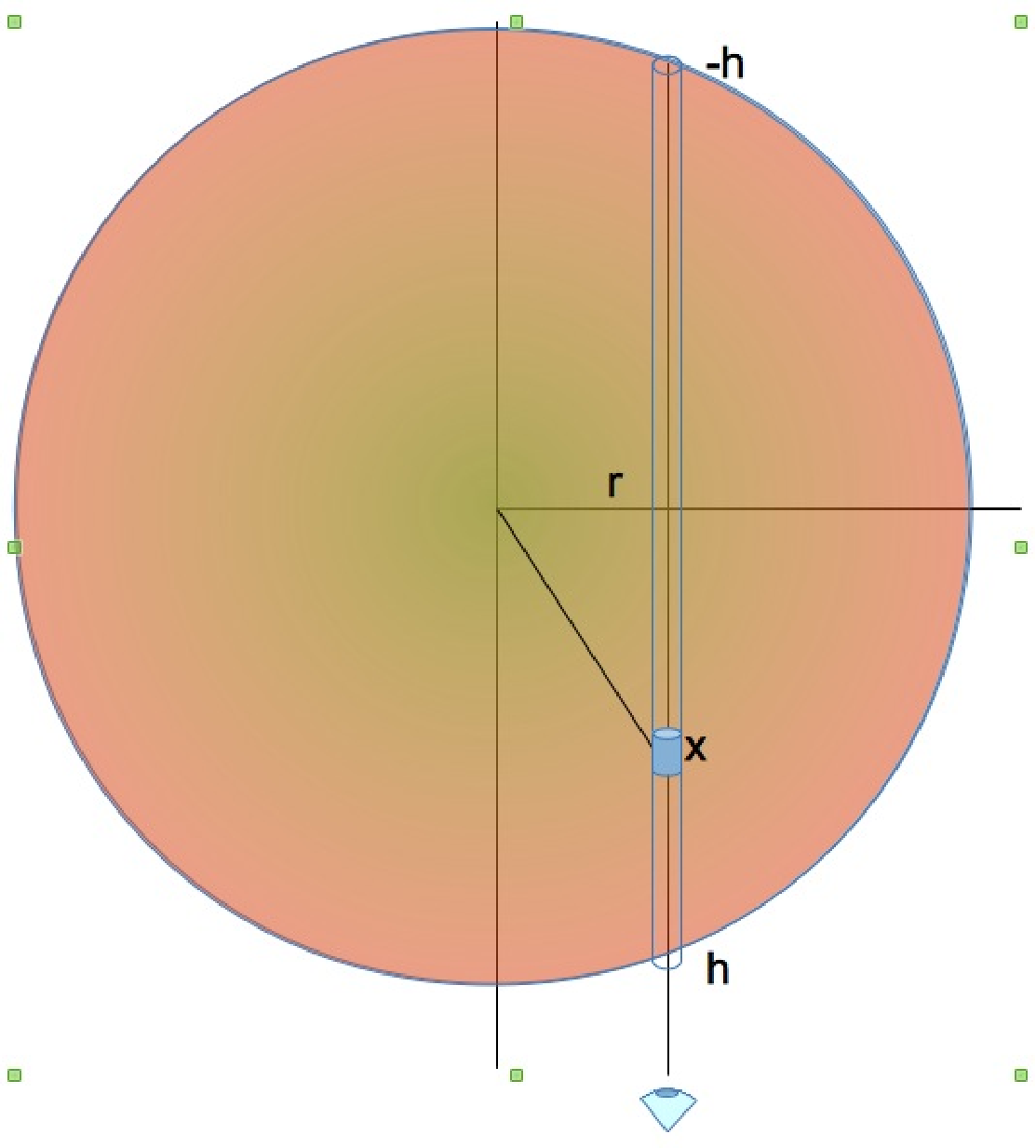}

  \caption{Sketch of the simple radiative transfer  model we built in order to simulate the differential H/K$_{s}$-band extinction effect  of a cuspy distribution of dust in which is embedded a cuspy distribution of stars.}
  \label{Fsketch}
\end{figure}
   
 At a given wavelength, the brightness at distance $r$ from the central peak
is given by
\begin{equation}
  F(r) \propto \int_{-h}^{h} L_{\star}(\lambda) \times  n_{\star}(\sqrt(r^{2}+x^{2})) \times  \exp(-\tau(x))  dx
 \end{equation} 
$n_{\star}(r)$ is the radial density of star, $ L_{\star}(\lambda)$ the star luminosity at $ \lambda$ and $ \tau(x)$  the optical depth between the observer and the cell at location $x$ on the line of sight. 
\begin{equation}
 \tau(x) =  \int_{-h}^{x} \alpha_{dust}(\sqrt(r^{2}+x^{2})) dx 
 \end{equation} 
  with  $\alpha_{dust}(r)$ the extinction coefficient of dust per unit of length at radius $r$. Indeed, $\alpha_{dust}$ depends on the wavelength, being proportional to $Q_{ext}(\lambda)$. 

%%%%%%%%%%%%%%%%%%%%%%%%%%%%%%%%
We developed a numerical code, in order to  look for the exponents of the cuspy distribution of  dust  and of the stars density that would give the best fit to the observed H-band  and K$_{s}$-band radial distributions. Using the Powell IDL procedure,  and leaving free both  exponents, we minimised the square of the log of the ratio of observed to simulated profiles, at H-band and K$_{s}$-band simultaneously. We set $\tau_{K_{s}} / \tau_{H} = 0.61$, i.e. the value found in the standard ISM. We reach a solution which is rather satisfactory, as shown on   Fig. \ref{model_fit1} where is plotted the radial fluxes at K$_{s}$-band and H,  modeled using a power-law for the stellar radial density of exponent   -4.2 and a  dust density varying as $r^{-1.05}$. On Fig. \ref{dust_cusp}, we plot the deduced radial distribution of the optical depth at H-band  and K$_{s}$-band down to the centre of the cusp: it appears indeed cuspy, as well.  From this plot, we can derive a radial optical depth $\tau_{H} = 7.7$ and $\tau_{K_{s}} = 4.7$ between 2.5 pc and 60 pc; this corresponds to   $\tau_{V} = 43$ in  V.  
%%%%%%%%%%%%%%%%%%%%%%%%%%%%%%%%%

This value would be close to twice the one derived by \cite{Burtscher16} and \cite{Burtscher15} who estimate $A_{v} = 24.6$. If our model is correct, the need for some extra extinction arising in the so-called torus would not be required, and the dusty cusp would have to be even truncated at  a radius of several pc or, another way to tell it, the torus could be identified with the tip of the dusty cusp. On the other hand,   \cite{Grosset18} using their  Monte-carlo radiative transfer code {\it Montagn} \citep{Grosset2016sf2a} to interpret AO-assisted near-IR polarisation measurements \citep{Gratadour2015}, estimate a minimum optical depth of 20 at K$_{s}$-band to the CE (see also next section). In that case,  the dusty cusp we suppose here would not be  sufficient to explain the most central extinction, and  a torus-like extra structure is still needed. Nevertheless, it is also possible that dichroism is involved too in the polarisation  seen at the centre, in which case a lower extinction would be required (Grosset, 2018, in preparation).  

If indeed the exponent we derive for the dust and stellar cusps are real, they require  to be explained on some theoretical basis. Especially, the exponent of -4 for the stellar density appears extremely steep and predicted in none of the theoretical studies we are aware of. We explore in the next subsection the possible role of a segregation of the stellar population with radius, where the intrinsic H-K$_{s}$ colour would increase when approaching the center.

 \begin{figure}[ht!]
 \centering    

 \includegraphics[width=0.50\textwidth,clip]{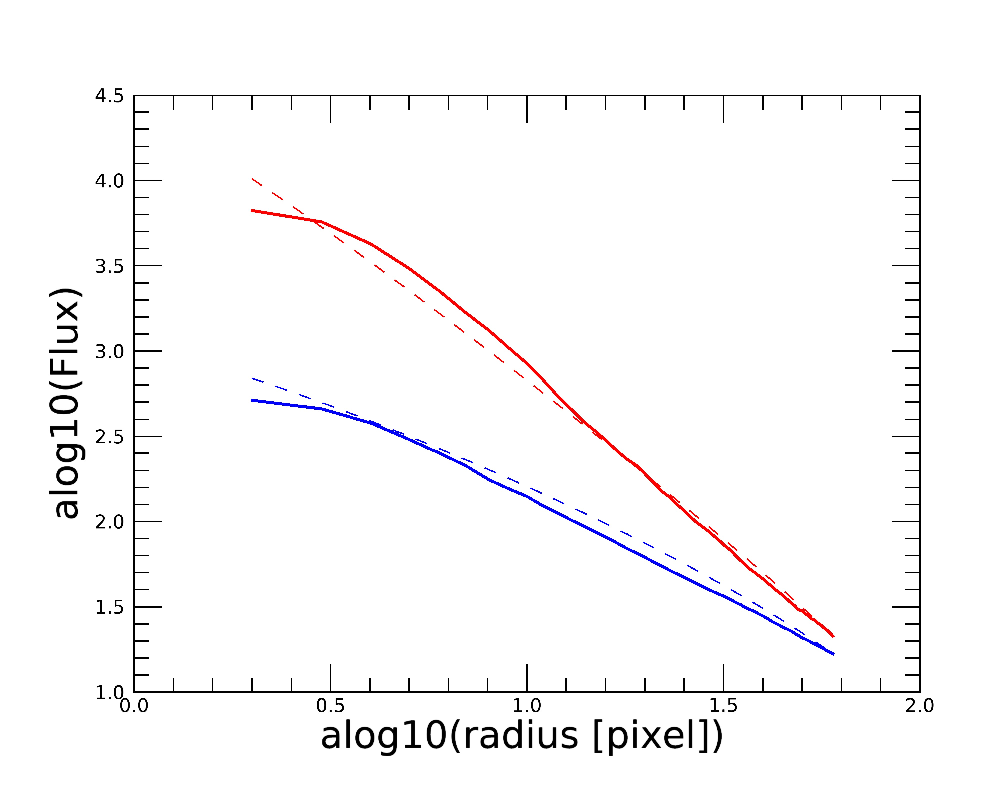}

  \caption{Red and blue solid lines: observed 2 to 50 pc  radial fluxes in K$_{s}$-band and H; red and blue dash lines:   
   K$_{s}$-band and H-band profiles resulting from our radiative transfer model, when the power-law for the stellar and dust radial density have exponents of respectively  -4.20  and -1.05. The result is  a quasi-power-law for both brightness profiles.   }
  \label{model_fit1}
\end{figure}

  \begin{figure}[ht!]
 \centering    
 \includegraphics[width=0.50\textwidth,clip]{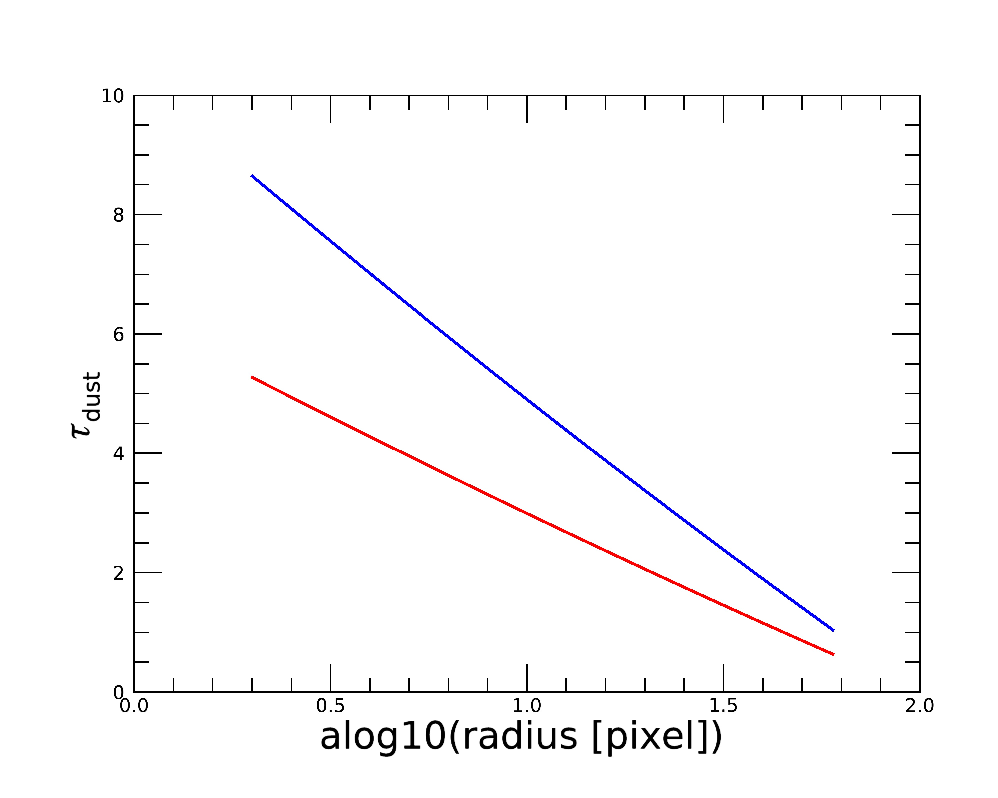} 
  \caption{Radial distribution of the  optical depth in H-band (blue) and K$_{s}$-band (red)  in the model providing the best fit to 
  the H-band and K$_{s}$-band observed brightness profiles (see text).  }
  \label{dust_cusp}
\end{figure}

 \subsection{Combination of  differential dust extinction and stellar type segregation}
 
We introduced in our numerical model the new constraint that the quantity H-K$_{s}$ varies
as the Log of the radius : H-K$_{s}$ = $ C \times log10(r/r_{max})/log10(r_{in}/r_{max}) $, assuming thus that  
H-K$_{s}$ = 0 at $r_{max} $ and H-K$_{s}$ = $C$ at $r_{in}$. This means that the un-redenned ratio  $Flux_{K_{s}}/Flux_{H}$  varies as a power-law vs $r$ as $ r^{C/2.5/log10(r_{in}/r_{max})}$. For instance, if $r_{in} = $ 1 pc and $r_{max} $ =  100 pc, the exponent of the power-law would be   $- C / 5$
On the other hand, the maximum value of $C$, assuming that the very central region is dominated by red giants is 0.3 (e.g. \cite{Ducati01}). One cannot thus expect a  power-law much steeper than  $ r^{-.06}$ and  the effect on the final fit, after introduction of this new ingredient in the model, cannot be to change significantly the exponent of the stellar cusp, nor of the dusty cusp. 

This was indeed confirmed by performing the fit that leads to  the values of exponents   -3.75 and -1.27 for respectively 
the stellar and the dusty cusps. Of course the stellar cusp exponent is now somewhat less extreme than the value -4.05 
we previously determined, but remains still very large  compared to any prediction done by numerical simulations or 
even observed in the well evidenced cusp of the Galactic Center \citep{Genzel03,Schodel17} where the exponent is -2.0 for 0.4 pc $< r < $4 pc. We  however note that the mass of the black hole SgrA* is significantly smaller than in NGC 1068. 

 \subsection{Colors and luminosity of the stellar cusp}
 
Up to now,  our aim has been essentially to reproduce the H-band and K$_{s}$-band  slopes of the radial profiles, 
without considering the measured fluxes themselves. Let's now look if there is some consistency 
between the estimated fluxes of the stellar cusp and the model we developed.  The measured  H-band 
and K$_{s}$-band magnitudes of the cusp solely, within a  1.''47 diameter  aperture (102 pc) centered on the 
peak,  are respectively    (9.51, 8.63)
Assuming that we correctly evaluated the central point-source magnitude in K$_{s}$-band (see below section 
\ref{Sec_PointSource}), we derive then for the supposed stellar cusp a magnitude Ks$_{cusp}$ = 8.9. This 
means an observed color (H-Ks)$_{cusp}$ = 0.6. Now, if we compute in our model, for each filter,  
the ratio  of the intrinsic  cusp  luminosity to the  one escaping from the dusty cusp, we can derive 
the corresponding  global (H-K$_{s}$) due to reddenning. It amounts to 1.6, assuming the second 
version of our model, i.e. combining  differential dust extinction and stellar type segregation.  The 
intrinsic  (H-Ks)$_{cusp}^{0}$ would then be -1.0. Even considering the uncertainties and the crudeness 
of the model,  this value appears  inconsistent with a stellar population. For instance, for a B1 star  
(H-K$_{s}$) = -0.14 \citep{Koornneef83} and a very young stellar cluster (H-K$_{s}$) = -0.1 -- 0.2 
\citep{Santos13}. This is a first clue that the hypothesis of pure stellar cusp is not robust enough.

Concerning now the apparent luminosity of the cusp at K$_{s}$-band it amounts to $L_{Ks} = 3.2 10^{8} \Lsun$. 
On the other hand, our model gives a fraction of escaping stellar light at K$_{s}$-band of 5.5 \%, leading to an 
intrinsic K$_{s}$-band luminosity of  $4.8 10^{9} 
\Lsun$. This is clearly  too high a value, since the ratio  $L_{bol}$ to $L_{Ks} $  is at least  20 for 
a cluster of 5 Gyr (e.g. see Fig. 5 of \cite{Thatte97}), leading to a total intrinsic stellar luminosity of 
the cusp of  $ > 1.~ 10^{11} \Lsun$, i.e. between 1/4  \citep{Raban09} and 2/3  \citep{Bock00} of the 
nuclear bolometric luminosity. This is probably the point which invalidates the interpretation 
in terms of pure stellar cusp:  this fraction  is much  larger than  usually accepted, especially if we 
consider that a cluster  5 Gyr old of this luminosity would have a mass of 2.8 $10^{11} ~\Msun$  
according to \cite{Thatte97} (Fig. 5). In addition, the involved quantity of dust  and thus of gas would
appear as well unrealistic. 

\section{Warm/hot dust extended emission}
\label{hot_dust}
We examine now if there is an alternative interpretation of the power-law behaviour of at least the K$_{s}$-band 
brightness radial profile in terms of thermal emission by warm to hot dust heated by the energetic 
radiation from the central engine. Of course this interpretation requires to be somewhat consistent 
with several known properties of the nuclear region. 

One first requirement is that the region where the thermal radiation dominates over other sources, 
especially stellar radiation, must extend at least to 50-60 pc, i.e. where the cuspy distribution is 
clearly observed. This requirement means that the dust temperature cannot drop rapidly along a 
radius, or in other terms, that the heating radiation from the CE cannot be attenuated on a 
small scale length. A very rough estimate of what this requirement means can be done  as follows. 
Let's consider  that the power-law in $r^{-2}$ is verified in the range 6 - 60 pc, thus that the ratio of thermal emission integrated on the line of sight between 6 and 60 pc should be 1/100.  Assuming that the 
medium is optically thin and homogeneous, the  dust luminosity per unit volume between 6 and 60 
pc must be in the ratio 1/1000, because the effective path  on which significant emission is integrated towards 
the observer is 10 times longer at 60 pc.  Assuming a radius of sublimation $R_{sub}$ = 2 pc, a 
temperature of dust sublimation $T_{sub} $ = 1500 K, and a dust coefficient of absorption  
Q$_{abs}(\lambda) \propto \lambda^{-2}$, the dust temperature would vary as $T(r) = 1500 (r/
R_{sub})^{-2/6}$ if the heating flux varies as $r^{-2}$, i.e. is not attenuated.  In that case, the ratio of 
dust luminosity per unit volume  between 6 and 60 pc would then be approximatively  1000 at 2.2 
\mic, as required. The hypothesis that the CE light is not attenuated  may look an unrealistic 
constraint because we know that, at least on the  direct line of sight to  the CE, there must be  an 
extinction  A$_{v} \approx 20$, attributed to the torus. However this does not  prevent that in the 
dust-free bicone, perpendicular to the torus, the extinction can be   low enough in a significant  fraction 
of the volume around the  CE which can thus be illuminated directly  by the energetic photons.  An 
alternative case could be that the medium is sufficiently clumpy so that  a fraction of direct light can 
indeed reach   regions distant from the CE and heat the local dust at a high enough temperature.  
Scattering on the surface of clumps should also help for energetic photons to travel up to those 
rather large distances. This is in fact the situation that we will  favour in the following.  To assess how realistic  this hypothesis is, we built a model of a populations of spherical absorbing/scattering clouds of various radii (uniformly distributed between 0 and 1.6 pc) within a sphere of radius 50 pc and launched 10$^{6}$ photons.  For a volume filling factor of 2.5\% and 300 clouds, the probability for a photon to reach the external radius of the sphere without being absorbed is 42 \%, Fig.  \ref{histo_clumpy_sphere} illustrates the histogram of radii reached by photons in such a simulation, where an albedo of 0.5 was taken.  One notes the very flat distribution of radii.

  \begin{figure}[ht!]
 \centering    
 \includegraphics[width=0.50\textwidth,clip]{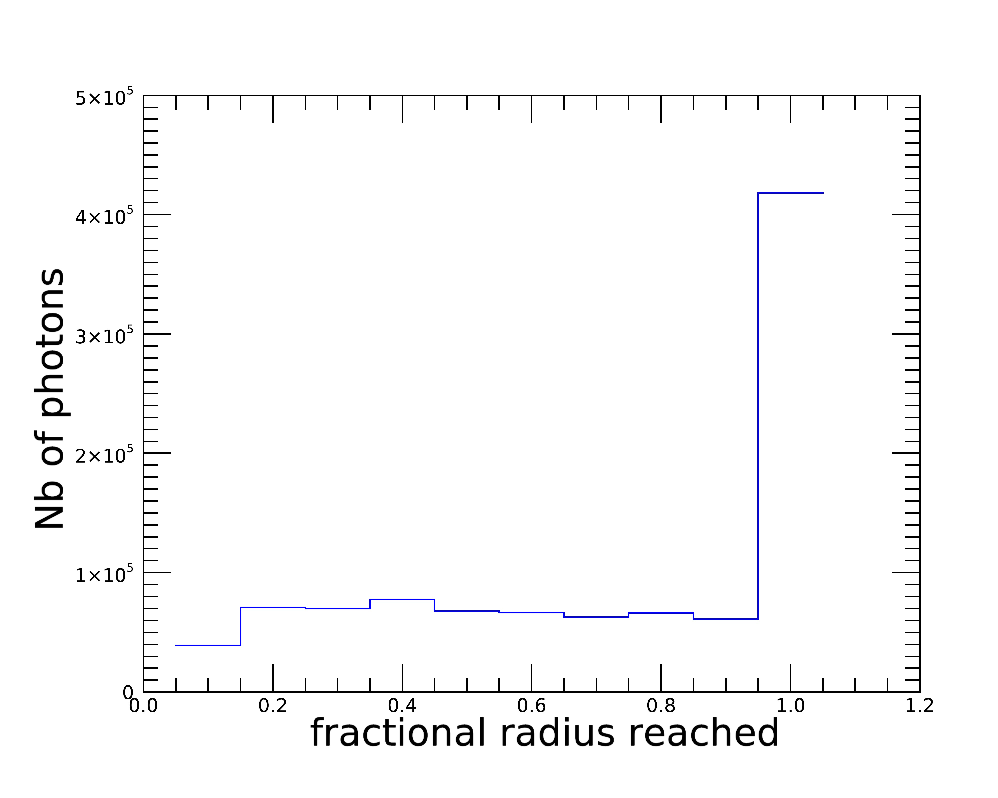} 
  \caption{Histogram of the radius reached by photons launched from the center of a sphere populated by 300 absorbing clouds totalling  a filling factor of 2.5\%.  The last and highest bin corresponds to escaping photons. Note the very flat distribution of radii. }
  \label{histo_clumpy_sphere}
\end{figure}
 
Assuming thus that dust can be heated significantly at least 50 pc away from the CE,   
we adapted the numerical code developed for the stellar cusp simulation (see section \ref{Sec_extinction}) to model  the situation described above. The luminosity of the dust per unit volume  is simply taken proportional to the Planck's law at the local temperature.  We also assume that there could be some extinction on the line of sight. Note that this last ingredient is not in contradiction with the idea that direct light from the CE can reach the dust at any distance, because this last assumption stems from the clumpiness of the medium, while the extended dust thermal emission could  
suffer extinction by  clumps along the line of sight. 

By adjusting very few parameters, namely a sublimation radius of 1pc, a constant dust density and $\tau_{Ks}$ = 6.5 from the center to 50 pc, we are able to fit fairly well  the measured K$_{s}$-band radial profile, as illustrated on Fig. \ref{fit_K_hot_dust} by the red dashed line. However, despite the quality of the fit,  the required K$_{s}$-band extinction is fairly large and
hardly compatible with our hypothesis of heating by  light directly from the CE. More over, when  examining  the case of the  H-band, assuming that,  here too,  dust thermal emission  is dominant, we find in fact no solution by our model consistent with the parameters found for the K$_{s}$-band fit. The least worse, indicated by the dotted blue line on Fig. \ref{fit_K_hot_dust} requires $\tau_{H}$ = 14.5 a value unconsistent with the classical grain extinction curve where $\tau_{Ks} / \tau_{H}$ = 0.61 (against 0.44 here) and moreover, no fit is found.  Especially, the temperature at large distance is too low to produce a significant flux in H.  

\begin{figure}[ht!]
 \centering    
 \includegraphics[width=0.50\textwidth,clip]{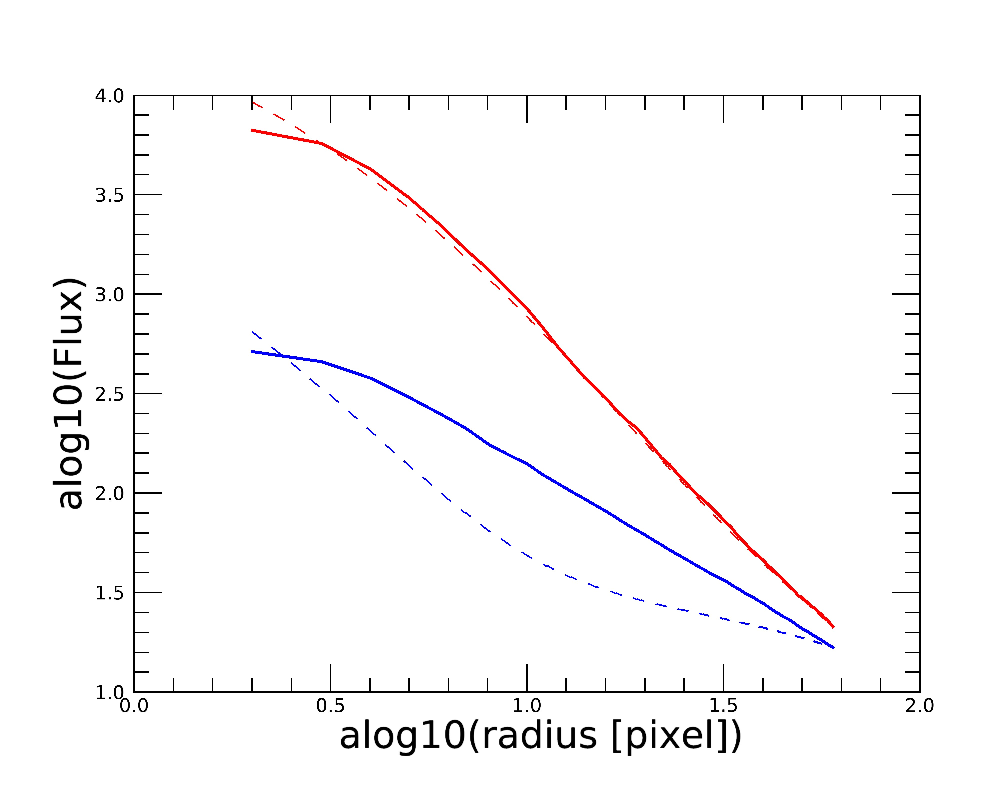} 
  \caption{K$_{s}$-band (red solid) and H-band (blue solid) radial profiles. Red dash line:  fit to the K$_{s}$-band profile assuming a
 maximum $\tau_{K}$ = 6.5 on the line of sight.  Blue dash line:   H-band profile of dust emission with  $\tau_{H}$ = 15 (see text).   }
  \label{fit_K_hot_dust}
\end{figure}

The unescapable conclusion is that we do not actually detect  dust thermal emission in H-band but only  light of a stellar cusp. This leads us to consider finally, rather than the pure K$_{s}$-band profile, the same one from which the stellar cusp contribution is subtracted, assuming that the later is just proportional to the H-band profile. We used a weighting factor 0.8 on the subtracted H-band profile. This corresponds to H-K$_{s}$ = 0.25, a value usually found in globular clusters .  We obtain the  green profile on Fig. \ref{fit_K-H_hot_dust}.  A pretty good fit by our model is found (green dashed line on Fig. \ref{fit_K-H_hot_dust}) with now  $\tau_{K_{s}} < 1.5$, a range quite consistent with the hypothesis of clumpiness.  We will actually assume   $\tau_{K_{s}} << 1$ in the following.

\begin{figure}[ht!]
 \centering    
 \includegraphics[width=0.50\textwidth,clip]{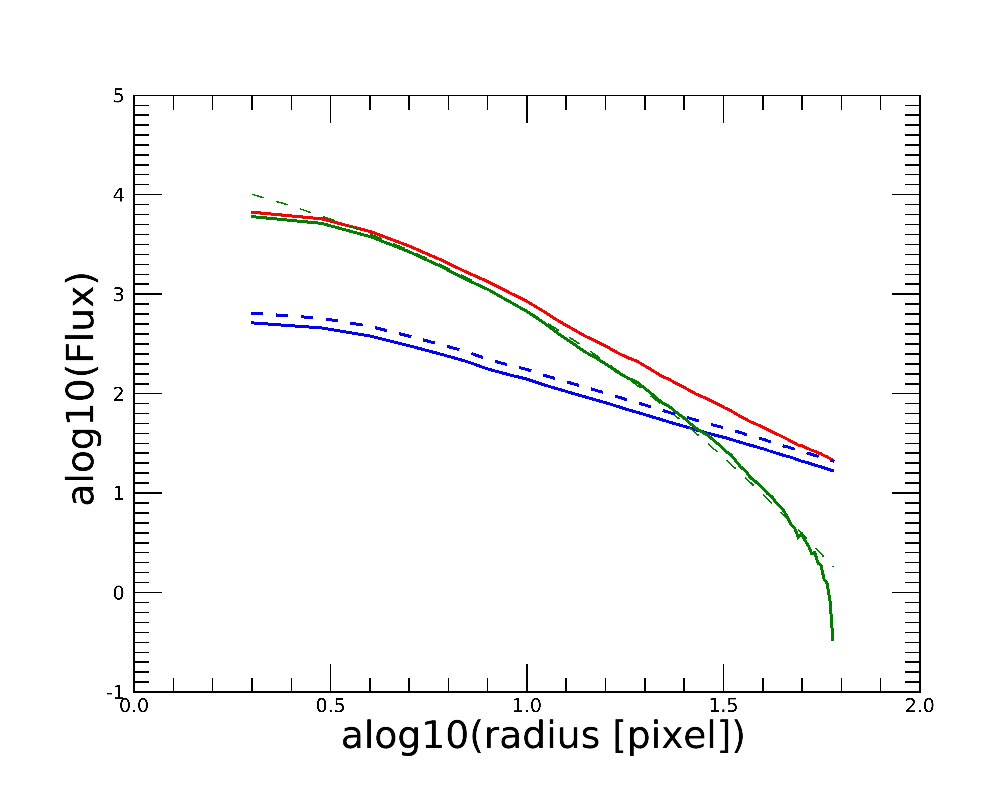} 
  \caption{K$_{s}$-band (red solid) and H-band (blue solid) radial profiles; green  solid line: K$_{s}$-band  radial profile subtracted with a stellar cusp contribution proportional to the H-band profile; blue dash line: estimated stellar cusp at K$_{s}$; green dash line: best fit by the model of thermal dust emission (see text).   }
  \label{fit_K-H_hot_dust}
\end{figure}
 
 Given the assumed low extinction, the K$_{s}$-band magnitude of the hot dust can be estimated  to be 9.55. 
 The spectral luminosity at K$_{s}$-band is then 5.0 $10^{34}$ W, that can be translated into $L_{bol} = 3.3~10^{37}$W, using our model. With the same model, we can relate this luminosity, derived from observation, to the content in warm dust, using some classical assumptions on the dust: MRN size distribution as $a^{-3.5}$, $a_{min} = 0.5 $nm,   $a_{max} = 500 $nm,  $Q_{abs}$ from \cite{Draine1985}. The  mass of dust  accounting for the observed luminosity, using the  temperature distribution of our model, is M$_{dust}$ = 0.07 \Msun, a value which  definitely corresponds to a very diluted medium. For instance, we derive a typical hydrogen density of about 2000 m$^{-3}$,  so that assuming  $\tau_{Ks} << 1$ is thus fully justified.     
 
 We think that with this final configuration we reached a fully consistent picture, where the nature of the two cuspy distributions in H-band and K$_{s}$-band are in fact largely different: it is totally dominated by a stellar cusp at H-band and, at K$_{s}$-band, it is  progressively dominated by warm to hot dust thermal radiation when going inward, this hot dust belonging to a very diluted medium.  We now can understand why the half-light radii at H-band and  K$_{s}$-band cusps, correspond to very different scales, as noted previously: the dust thermal emission dominating the central region in K$_{s}$,  is indeed much more compact than the stellar cusp seen in H.
 
\section{The nature of the central point like source}
\label{Sec_PointSource}
As shown in section\ref{sec-data}, the K$_{s}$-band radial profile reveals that there is a quasi-pointlike source 
superimposed on the power-law aisles  discussed above. Using the SPHERE calibration provided by 
ESO and estimating  the best PSF fitting to the point source, we find m$Ks_{point} = 10.2 \pm 0.2$. 
The point-source provides 31\% of the K$_{s}$-band luminosity measured in the 50 pc diameter central part. 
This number is significantly   smaller than estimated by \cite{Thatte97}, from their speckle and 
spectroscopic data, albeit  at a lower spatial resolution. 

We estimated that the size of the source cannot be larger that 1.7 pc.  As proposed by several authors \citep{Thatte97, Gratadour2003}, a straightforward interpretation is that the source corresponds to the emission by dust at high temperature, close to sublimation, at the internal walls of the torus.  We considered various estimates of this internal radius : \cite{Nenkova2002} propose the relation $R_{in} =  1.2~ L_{12}^{1/2}$ ~pc, with $ L_{12} = 10^{-12}~L_{bol}/\Lsun$  which leads to  $R_{in} =  0.9~ $pc, using $L_{bol} =  2.1~10^{45} ~erg/s$  \citep{Marconi2004}; with $L_{bol} =  4.2~10^{44} ~erg/s$ as estimated by \cite{Garcia-Burillo2014} we rather find $R_{in} =  0.4~ $pc. This is to be compared to the maximum radius of 0.8 pc we mentioned in section \ref{sec-data} for the central point source: there is thus a good consistency with  the picture suggested by our SPHERE data. 

As an additional step, we can compare  the flux we measure for the point source  to some estimate of the power emitted by the assumed hot dust. At the distance of NGC 1068, K$_{s}$-band = 10.2 means a K$_{s}$-band spectral luminosity of  $2.9~ 10^{34}$ W ~$\mic^{-1}$. Assuming a sublimation temperature of 1400 K and a central cavity radius of 0.6 pc  with walls optically thick at 2.2~\mic, the expected intrinsic K$_{s}$-band spectral luminosity from the cavity would be $\approx \pi~ B_{\lambda} ~ 2 \pi~ R_{in}^{2}  =  4.3~10^{37}$ W $\mic^{-1}$ (assuming an opening angle of the bicone of 120\dgr). This means that  7.9 magnitude of extinction in K$_{s}$-band is required to reconcile the two numbers.   This is significantly lower than the 20 mag in K$_{s}$-band that  \cite{Grosset18} propose to explain their near-IR polarisation data, but somewhat larger than   the range 2.5 -- 4.8 of  extinction  proposed by several authors \citep{Thatte97,Efstathiou95,Young95}. Given our  rough estimate of the K$_{s}$-band luminosity of the very host dust, that is based on an approximate  $ R_{in}$ and geometry,  neglects any radiative transfer effect, including scattering, nor considers a realistic dust composition, the uncertainty is probably larger than 2.5 magnitudes. In any cases, the essential of the obscuration must occur in the very central part, thus supporting the hypothesis of a  torus with a dense internal shell at a pc scale. This does not prevent, that a more extended molecular extension of the torus  exists on a 7-10 pc scale, as the one identified by \cite{Garcia-Burillo16}, and even a more diluted envelope on a 30 pc scale, as proposed by \cite{Gratadour2015} to explain the peculiar near-IR polarisation pattern.

\section{Conclusion and Prospectives}

The high angular resolution near-IR images of the central 200 pc  of of NGC 1068 reveal a cuspy distribution of the brightness at all wavelengths between H-band and K$_{s}$-band with, however, a continuously increasing absolute value of the exponent with wavelength. Using a simple numerical  model, we tried first to interpret this behaviour as resulting from a very cuspy  ( $r^{-4}$) distribution of the stellar density in a central compact cluster down to 1 pc, while the H/K$_{s}$-band cusp's exponent difference  would be  accounted for by a distribution of the ISM as well cuspy, but with a less steep exponent $\approx$ -1.  Introducing some segregation in the stellar population, i.e. assuming that giant stars are dominating towards the very center, changes only slightly  the exponent  of the stellar cusp that becomes -3.75, i.e. a value never observed, nor predicted on theoretical grounds. The required dust opacity is in all cases very high, so that, when computing the intrinsic luminosity of  the putative stellar cluster, we  reach a value much too high to be  realistic.   

The alternative interpretation that we explored is that the K$_{s}$-band brightness profile is a combination of  
a stellar cusp and  warm to hot dust emission and it appears clearly more acceptable. Considering the residual profile when a stellar cusp component, derived from the H-band profile, is subtracted, a good fit  can be obtained by a simple model of dust heated by the CE in a homogeneous and diluted medium. 

In addition to the cusp and thermal warm dust radiation, the K$_{s}$-band profile features a distinct  central point-like source interpreted as the very hot dust, close to the central engine. The estimated extinction to the very  center is consistent with previous estimates. The   overall picture revealed by this study appears well consistent with the  scheme of a clumpy medium around the CE and with  the requirement of a torus at a pc scale.  It also brings  constraints on the structure of
the ISM within the 100 central pc, which must be very diluted.   

As a final remark, we note that the three components we identified within the central 100 pc, namely the stellar cusp, the very hot dust core and the extended warm dust, contribute each to the K$_{s}$-band luminosity for  a similar fraction, with magnitudes in K$_{s}$-band respectively of 9.8, 10.2 and 9.6.    

Several questions are opened by this study,  let us mention at least few of them:  can a similar  stellar cusp and extended dust emission be observable in other AGNs thanks to HAR imaging or even interferometry ? Can  radiative transfer models produce a consistent view of all observations, including recent ALMA molecular detection of the torus, mid-IR imaging, mid-IR interferometry  and  near-IR polarisation?  Can   the rather steep stellar cusp   be predicted by theoretical models combining multi-scale hydrodynamics, stellar dynamics and star formation ? Is the relationship that we established from the data of \cite{Haan2013}, between cusp radius, cusp luminosity and exponent verified for other cases of nearby AGNs ? Is there a solid explanation for this phenomenological law, especially that the exponent depends only on the ratio $L_{cusp} / R_{cusp}$  ? In any case, it is in the views of our team to tackle the two first questions, especially  the second by using our numerical model {\it Montagn}.    
% Optional acknowledgements
% -------------------------
\begin{acknowledgements}
The authors would like to acknowledge financial support from the Programme National Hautes Energies (PNHE) and from "Programme National de Cosmologie and
Galaxies" (PNCG) funded by CNRS/INSU-IN2P3-INP, CEA and CNES, France.
\end{acknowledgements}


\bibliographystyle{aa} % style aa.bst
\bibliography{rouan.bib} % your references Yourfile.bib

%\begin{appendix} %First appendix

%\label{pelat}
%\end{appendix} 

\end{document}